\documentclass[structabstract]{aa}  
\usepackage{graphicx}
\usepackage{txfonts}
\usepackage{natbib}
\usepackage[breaklinks=false]{hyperref}
\newcommand\dif{{\rm d}}
\newcommand\dln{{\rm dln}}

\newcommand\Ms{{\ensuremath{\mathrm{M}_{\odot}}}}

\newcommand\Ls{{\ensuremath{\mathrm{L}_{\odot}}}}

\newcommand\Mpy{\Ms\,{\rm yr}{\ensuremath{^{-1}}}}

\newcommand\dm{\ensuremath{\dot M}}

\newcommand\dphi{{\ensuremath{\dif\varphi}}}
\newcommand{\genec}{{\sc genec}}
\newcommand{\dt}{{\ensuremath{\dif t}}}
\newcommand{\dr}{{\ensuremath{\dif r}}}
\newcommand{\ds}{{\ensuremath{\dif s}}}
\newcommand{\dth}{{\ensuremath{\dif\theta}}}
\newcommand\dMr{{\ensuremath{\dif M_r}}}

\begin{document}

\title{The properties of supermassive stars in galaxy merger driven direct collapse I:
models without rotation}
\titlerunning{The properties of supermassive stars in galaxy merger driven direct collapse I}

\author{
L. Haemmerl\'e
}
\authorrunning{Haemmerl\'e}

\institute{D\'epartement d'Astronomie, Universit\'e de Gen\`eve, chemin Pegasi 51, CH-1290 Versoix, Switzerland}


 
\abstract
{The formation of the most massive quasars observed at high redshifts requires extreme accretion rates ($>1$ \Mpy).
Inflows of 10 -- 1000 \Mpy\ are found in hydrodynamical simulations of galaxy mergers,
leading to the formation of supermassive discs (SMDs) with high metallicities ($>$ Z$_\odot$).
Supermassive stars (SMSs) born in these SMDs could be the progenitors of the most extreme quasars.}
{Here, we study the properties of non-rotating SMSs forming in high metallicity SMDs.}
{Using the stellar evolution code \genec, we compute numerically the hydrostatic structures of non-rotating SMSs
with metallicities $Z=1-10$ Z$_\odot$ by following their evolution under constant accretion at rates 10 -- 1000 \Mpy.
We determine the final mass of the SMSs, set by the general-relativistic (GR) instability,
by applying the relativistic equation of adiabatic pulsations to the hydrostatic structures.}
{We find that non-rotating SMSs with metallicities $Z=1-10$ Z$_\odot$ accreting at rates 10 -- 1000 \Mpy\
evolve as red supergiant protostars until their final collapse.
All the models reach the GR instability during H-burning.
The final mass is $\sim10^6$~\Ms, nearly independently of the metallicity and the accretion rate.}
{}

   \keywords{stars: massive -- stars: black holes}
 
\maketitle
%

\section{Introduction}
\label{sec-in}

In the last decade, a number of quasars, hosting supermassive black holes (SMBHs) with masses $10^6-10^9$~\Ms,
have been discovered at redshifts $z\gtrsim7$
\citep{wu2015,banados2018,wang2018,yang2020,wang2021a,larson2023,kokorev2023,kovacs2024,bogdan2024,maiolino2024}.
The young age of the universe at these redshifts implies an extremely rapid formation process for these SMBHs
\citep{woods2019,haemmerle2020a}.
A promising scenario is direct collapse, that is the direct formation of a SMBH seed with mass $\gtrsim10^5-10^6$ \Ms.
In this scenario, the progenitor of the SMBH seed is a supermassive star (SMS) growing by accretion
\citep{hosokawa2012a,hosokawa2013,umeda2016,woods2017,woods2021a,woods2024,haemmerle2018a,haemmerle2018b,haemmerle2019c,herrington2023,nandal2023,nandal2024}
until it reaches the general-relativistic (GR) instability
\citep{chandrasekhar1964,haemmerle2020c,haemmerle2021a,haemmerle2021b,haemmerle2024,saio2024,nagele2024}.

Direct collapse requires special conditions.
In the collapse of an isolated primordial halo, molecular cooling by H$_2$ is found to limit the sellar mass
to $\sim100$ \Ms\ (e.g.~\citealt{bromm2002a,klessen2023}).
Only in the presence of a strong Lyman-Werner flux, that dissociates H$_2$,
a zero-metallicity SMS can form, under accretion at rates 0.01 -- 1 \Mpy, in the collapse of the atomically cooled halo
(e.g.~\citealt{haiman1997a,omukai2001a,dijkstra2008,bromm2003b,latif2013e,regan2017,chon2018}).
At a redshift $z\sim7$, the age of the universe is a fraction of a Gyr,
so that accretion rates $>1$~\Mpy are required to form objects like
ULAS J1120+0641 \citep{mortlock2011}, ULAS J1342+0928 \citep{banados2018},
DELS J0038-1527 \citep{wang2018}, UHS J1007+2115 \citep{yang2020} or J0313-1806 \citep{wang2021a},
hosting SMBHs with masses $\sim10^9$ \Ms.
We see that the rates of atomically cooled halos are just at the limit of what is required
to form the most massive quasars observed at high redshifts.

A more efficient route of direct collapse is the case of the merger of massive gas-rich galaxies
\citep{mayer2010,mayer2015,mayer2024,zwick2023}.
In this scenario, inflows of 10~--~1000~\Mpy\ are triggered down to parsec scales
(see Figure~5 of \citealt{mayer2024}),
which results in the formation of a rotationally supported supermassive disc (SMD) with mass $\sim10^9$~\Ms.
Cosmological simulations indicate that such SMDs are metal-rich, with a metallicity in the range 1 -- 10 Z$_\odot$,
typically 3 Z$_\odot$ (see Figure~7 of \citealt{mayer2024}).

Here, we study for the first time the properties of SMSs forming by accretion
in the conditions of galaxy merger driven direct collapse,
that is with metallicities $>$ Z$_\odot$ and accretion rates 10~--~1000~\Mpy.
We derive numerically the properties of such SMSs during their hydrostatic accretion phase
as well as their final masses at collapse, set by the GR instability,
which provides an estimate of the mass of the SMBH seeds formed in this scenario.
This work is the first of a series, and here we consider only the non-rotating case.
The method is described in Sect.~\ref{sec-meth},
the results are presented in Sect.~\ref{sec-res} and discussed in Sect.~\ref{sec-dis},
and we conclude in Sect.~\ref{sec-out}.

\section{Method}
\label{sec-meth}

\subsection{Hydrostatic structures}
\label{sec-genec}

The hydrostatic structures are computed with \genec\ \citep{eggenberger2008},
a one-dimensional hydrostatic stellar evolution code that solves numerically the equations of stellar structure
with the Henyey method.
The code includes the GR Tolman-Oppenheimer-Volkoff corrections to hydrostatic equilibrium
in the post-Newtonian approximation \citep{haemmerle2018a}.
Accretion is included under the assumption of cold accretion \citep{haemmerle2016a}.
We start from an initial model of $M=100$ \Ms, and consider constant accretion at rates $\dm=10-100-1000$ \Mpy.
We use such a high initial mass because the rates considered are not consistent with gravity for lower masses
\citep{haemmerle2021c}.

The chemical composition of the initial model and that of the accreted material are identical
and are scaled with the solar abundances of \cite{asplund2005}.
For each metallicity $Z=1-3-10$ Z$_\odot$, the mass fraction of hydrogen ($X$) and of helium ($Y$)
are set by extrapolation from the primordial and solar compositions:
\begin{equation}
Y=Y_0+\left(Y_\odot-Y_0\right){Z\over Z_\odot}
\end{equation}
\begin{equation}
X=1-Y-Z
\end{equation}
where $Y_0=0.2484$ is the primordial helium mass fraction.

\subsection{GR instability}
\label{sec-gr}

We determine the GR instability point with the method of \cite{chandrasekhar1964}.
This method relies on a linear pulsation analysis of Einstein's field equations in spherical symmetry,
assuming adiabatic perturbations to equilibrium.
The pulsation equation reads
\begin{eqnarray}
{\omega^2\over c^2}\cdot r^2e^{-a}\xi(r)&=&
{r^2e^{-a-3b}\over P+\rho c^2}\cdot\left(
-\left(\Gamma_1P\,{e^{3a+b}\over r^2}\left(r^2e^{-a}\xi(r)\right)'\right)'
\right.\label{eq-radial6}\\&&\left.
+{e^{3a+b}\over r^2}\left({4\over r}P'-{P'^2\over P+\rho c^2}
\right.\right.\nonumber\\&&\left.\left.
+{8\pi G\over c^4}P\left(P+\rho c^2\right)e^{2b}\right)\cdot r^2e^{-a}\xi(r)\right)
\nonumber\end{eqnarray}
where $'$ denotes a derivative with respect to the radial coordinate $r$,
$\omega$ is the pulsation frequency, $\xi$ the radial displacement of the perturbation,
$\Gamma_1$ the first adiabatic exponent, $P$ the thermal pressure, $\rho$ the relativistic mass density,
$c$ the speed of light, $G$ the gravitational constant, and $a$ and $b$ are the coefficients of the metric:
\begin{equation}
\ds^2=-e^{2a}(c\dt)^2+e^{2b}\dr^2+r^2\dth^2+r^2\sin^2\theta\ \dphi^2
\label{eq-ds}\end{equation}
Except $\omega$ and $\xi$, all the quantities involved in equation (\ref{eq-radial6}) refer to equilibrium.

Equation~(\ref{eq-radial6}) is an eigenvalue problem
\begin{equation}
{\omega^2\over c^2}f=\mathcal{L}f
\end{equation}
for the Sturm-Liouville operator
\begin{eqnarray}
\mathcal{L}f(r)&=&
{r^2e^{-a-3b}\over P+\rho c^2}\cdot\left(-\left(\Gamma_1P\,{e^{3a+b}\over r^2}f'(r)\right)'
\right.\label{eq-sturm}\\&&\left.
+{e^{3a+b}\over r^2}\left({4\over r}P'-{P'^2\over P+\rho c^2}
+{8\pi G\over c^4}P\left(P+\rho c^2\right)e^{2b}\right)\cdot f(r)\right)
\nonumber\end{eqnarray}
with $f(r)=r^2e^{-a}\xi(r)$ and the boundary conditions $\xi(0)=0$ and $\delta P(R)=0$.
The operator $\mathcal{L}$ is self-adjoint for the scalar product
\begin{equation}
\langle f\vert g\rangle:=\int_0^R{P+\rho c^2\over r^2}\,e^{a+3b}f(r)g(r)\dr
\label{eq-scalaire}\end{equation}
As a consequence, the spectral theorem implies that there exists an orthonormal basis
(of the real vector space of functions satisfying the required boundary conditions)
whose elements are the eigenvectors of $\mathcal{L}$.
In other words, any function $f(r)$ that satisfies the required boundary conditions
can be written as a linear combination of eigenfunctions $f_i$ with eigenvalue $\lambda_i$:
\begin{equation}
f=\sum_ic_if_i
\end{equation}
And the different $f_i$ are all orthonormal to each other.
It implies that
\begin{equation}
\langle f\vert\mathcal{L}f\rangle=\sum_{i,j}c_ic_j\langle f_i\vert\mathcal{L}f_j\rangle
=\sum_{i,j}c_ic_j\lambda_j\underbrace{\langle f_i\vert f_j\rangle}_{=\delta_{ij}}
=\sum_ic_i^2\lambda_i
\label{eq-spectral}\end{equation}
We see that, if there is any function $f(r)$ whose scalar product with its image through $\mathcal{L}$ is negative,
then there exists necessarily at least one negative eigenvalue $\lambda_i<0$.
Indeed, if the left-hand side of Equation~(\ref{eq-spectral}) is negative, the right-hand side must also be negative,
which is possible only if one of the $\lambda_i$ is negative.
And if an eigenvalue of $\mathcal{L}$ is negative, it implies that there are solutions to Equation~(\ref{eq-radial6})
with an imaginary pulsation frequency, that is the star is GR unstable.

By applying the scalar product $\langle f\vert\cdot\rangle$ with $f(r)=r^2e^{-a}\xi(r)$
on the two sides of Equation (\ref{eq-radial6}), we obtain
\begin{eqnarray}
\omega^2\int_0^Re^{-a+3b}\left(1+{P\over\rho c^2}\right)\rho r^2\xi^2(r)\dr
\qquad\qquad\qquad\qquad\nonumber\\\qquad\qquad=
-\int_0^Rr^2e^{-a}\xi(r)\left(\Gamma_1P\,{e^{3a+b}\over r^2}\left(r^2e^{-a}\xi(r)\right)'\right)'\dr
\nonumber\\\qquad\qquad
+4\int_0^RP're^{a+b}\xi^2(r)\dr-\int_0^R{P'^2r^2e^{a+b}\over P+\rho c^2}\xi^2(r)\dr	\nonumber\\\qquad\qquad
+{8\pi G\over c^4}\int_0^R\left(P+\rho c^2\right)Pr^2e^{a+3b}\xi^2(r)\dr
\label{eq-radial7}\end{eqnarray}
We can transform the first term on the right-hand side by an integration by parts (the boundary terms vanish):
\begin{eqnarray}
\omega^2\int_0^Re^{-a+3b}\left(1+{P\over\rho c^2}\right)\rho r^2\xi^2(r)\dr
\qquad\qquad\qquad\qquad\qquad\nonumber\\=
\int_0^R\Gamma_1P\,{e^{3a+b}\over r^2}\left(r^2e^{-a}\xi(r)\right)'^2\dr
\qquad\qquad\qquad\qquad\nonumber\\\qquad\qquad\qquad
+4\int_0^RP're^{a+b}\xi^2(r)\dr
-\int_0^R{P'^2r^2e^{a+b}\over P+\rho c^2}\xi^2(r)\dr	\nonumber\\
+{8\pi G\over c^4}\int_0^R\left(P+\rho c^2\right)Pr^2e^{a+3b}\xi^2(r)\dr		\qquad\qquad
\label{eq-radial8}\end{eqnarray}
This is Equation (61) of \cite{chandrasekhar1964}.
It corresponds to
\begin{equation}
{\omega^2\over c^2}\langle f\vert f\rangle=\langle f\vert\mathcal{L}f\rangle
\end{equation}
And as we saw in Equation (\ref{eq-spectral}),
'{\it it is clear that a sufficient condition for the dynamical instability of a mass is that the right-hand side
of [Equation (\ref{eq-radial8})] vanishes for some chosen "trial function" $\xi$
which satisfies the required boundary conditions}' \citep{chandrasekhar1964}.
The function $\xi(r)$ does not need to represent the actual perturbation to equilibrium,
this is just an arbitrary trial function that allows to prove the existence of a negative eigenvalue.
For $\xi(r)=re^{a(r)}$, Equation (\ref{eq-radial8}) gives Equation (1) of \cite{haemmerle2021a}:
\begin{eqnarray}
{\omega^2\over c^2}I_0=\sum_{i=1}^4I_i
\label{eq-chandra}\end{eqnarray}
with
\begin{eqnarray}
I_0&=&\int_0^Re^{a+3b}(P+\rho c^2)r^4\dr					\label{eq-I0}\\
I_1&=&9\int_0^R\Gamma_1e^{3a+b}Pr^2\dr					\label{eq-I1}\\
I_2&=&4\int_0^Re^{3a+b}P'r^3\dr							\label{eq-I2}\\
I_3&=&{8\pi G\over c^4}\int_0^Re^{3(a+b)}P(P+\rho c^2)r^4\dr		\label{eq-I3}\\
I_4&=&-\int_0^Re^{3a+b}{P'^2\over P+\rho c^2}r^4\dr			\label{eq-I4}
\end{eqnarray}
The sufficient condition for instability obtained in this way
is more efficient than GR hydrodynamical stellar evolution codes
in capturing the GR instability \citep{haemmerle2021a,nagele2022}.
Moreover, the trial function used approximates accurately the actual dynamical perturbation,
which implies that this condition is close to an exact condition \citep{saio2024}.

In the post-Newtonian limit, Equation~(\ref{eq-chandra}) reduces to \citep{haemmerle2021b}
\begin{eqnarray}
\omega^2I&=&\int\beta P\dif V-\int\left({2GM_r\over rc^2}+{8\over3}{P\over\rho c^2}\right){GM_r\over r}\dMr
\label{eq-postnewton2}\end{eqnarray}
where $\beta$ is the ratio of gas pressure to total pressure, $\dif V=4\pi r^2\dr$, $M_r$ is the Lagrangian coordinate
and $I$ is the moment of inertia.

\section{Results}
\label{sec-res}

\begin{figure}\begin{center}
\includegraphics[width=.5\textwidth]{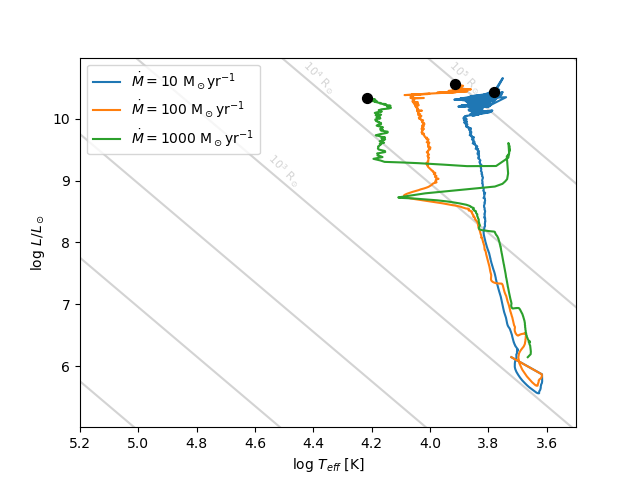}
\caption{HR diagram of the models at $Z=3$ Z$_\odot$.
The black circles indicate the point of GR instability.}
\label{fig-hrZ3}
\end{center}\end{figure}

The evolutionary tracks on the Hertzsprung-Russell (HR) diagram of the models at $Z=3$ Z$_\odot$
are shown in Figure~\ref{fig-hrZ3}.
The luminosity $L$ grows continuously with the stellar mass $M$, following nearly the Eddington limit:
\begin{equation}
L\simeq{4\pi cGM\over\kappa}
\end{equation}
where $\kappa\simeq0.2(1+X)$ cm$^2$ g$^{-1}$ is the opacity, dominated by electron scattering.
The effective temperature shows strong oscillations due to the low resolution in the outter layers,
but the star remains a 'red supergiant protostar' \citep{hosokawa2012a,hosokawa2013,haemmerle2018a}
all along the evolution, following the Hayashi limit up to luminosities $10^{10}-10^{11}$ \Ls.

\begin{figure}\begin{center}
\includegraphics[width=.5\textwidth]{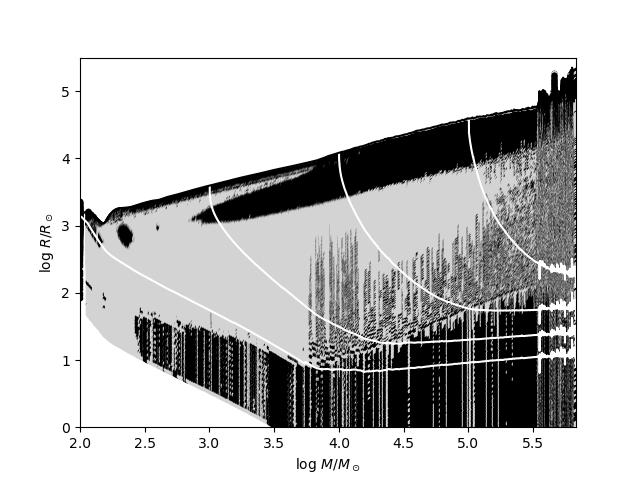}
\caption{Internal structure of the model at $Z=3$ Z$_\odot$ and $\dm=10$~\Mpy,
as a function of the stellar mass $M$.
The convective zones are shown in black and the radiative zones in grey.
The white curves are iso-masses (Lagrangian layers).}
\label{fig-stZ3_dm10}
\end{center}\end{figure}

\begin{figure}\begin{center}
\includegraphics[width=.5\textwidth]{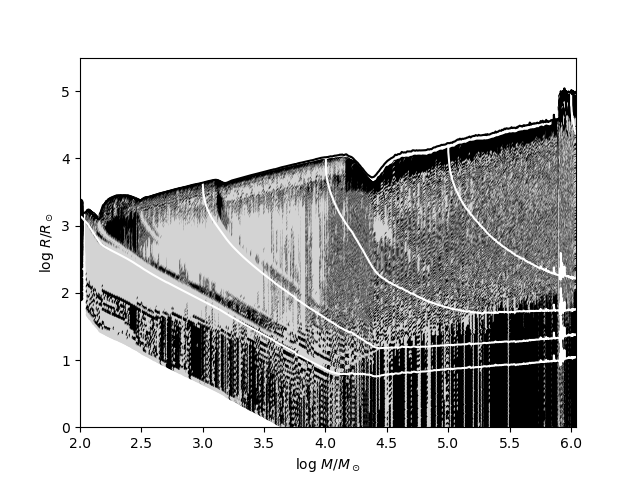}
\caption{Same as Figure~\ref{fig-stZ3_dm10} for the model at $Z=3$ Z$_\odot$ and $\dm=100$~\Mpy.}
\label{fig-stZ3_dm100}
\end{center}\end{figure}

\begin{figure}\begin{center}
\includegraphics[width=.5\textwidth]{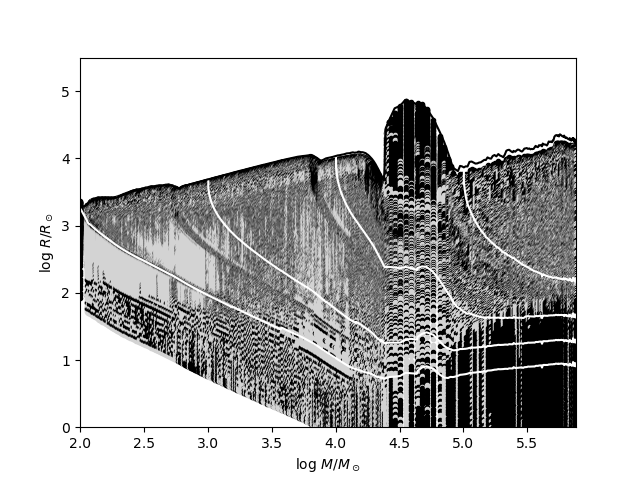}
\caption{Same as Figure~\ref{fig-stZ3_dm10} for the model at $Z=3$ Z$_\odot$ and $\dm=1000$~\Mpy.}
\label{fig-stZ3_dm1000}
\end{center}\end{figure}

\begin{figure}\begin{center}
\includegraphics[width=.5\textwidth]{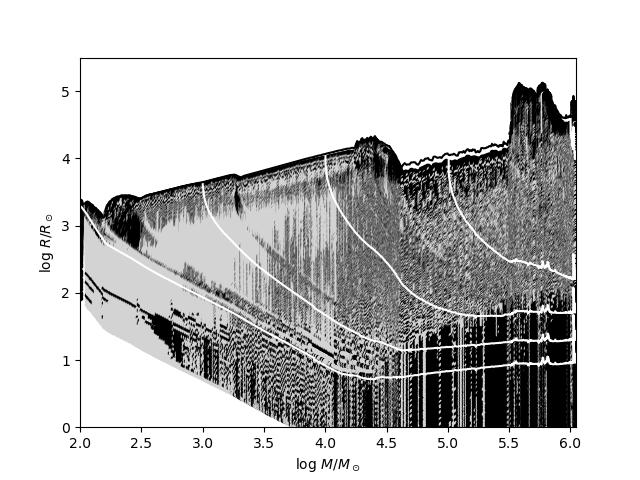}
\caption{Same as Figure~\ref{fig-stZ3_dm10} for the model at $Z=$ Z$_\odot$ and $\dm=100$~\Mpy.}
\label{fig-stZ1_dm100}
\end{center}\end{figure}

The internal structure of some of the models
are shown in Figures~\ref{fig-stZ3_dm10}-\ref{fig-stZ3_dm100}-\ref{fig-stZ3_dm1000}-\ref{fig-stZ1_dm100}.
Once the star is supermassive ($M\gtrsim10^4-10^5$~\Ms), it is made of a convective core,
where convection is driven by H-burning, a 'radiative' zone (full of small transient convective zones),
and a convective envelope due to the low temperatures and high opacities on the Hayashi limit.
The convective envelope is thiner for larger accretion rates.
Although the photospheric radius keeps growing, all the Lagrangian layers are contracting,
except in the convective core, where the heat liberated by H-burning causes the gas to expand.
We see again the oscillations of the surface already visible in Figure~\ref{fig-hrZ3}.
Each expansion of the photospheric radius results in the deepening of the convective envelope.
Once it enters the radiative region, the gas accreted during such events keeps a memory of it,
showing a higher density of transient convective zones compared to the other layers
(see for instance the layers $10^4$ \Ms $<M_r<10^5$~\Ms\ in Figure~\ref{fig-stZ3_dm1000}).
Inversely, the layers accreted during a contraction show a lower density of transient convective zones
(see for instance the layers $10^4$ \Ms $<M_r<10^5$ \Ms\ in Figure~\ref{fig-stZ3_dm100}).

\begin{figure}\begin{center}
\includegraphics[width=.5\textwidth]{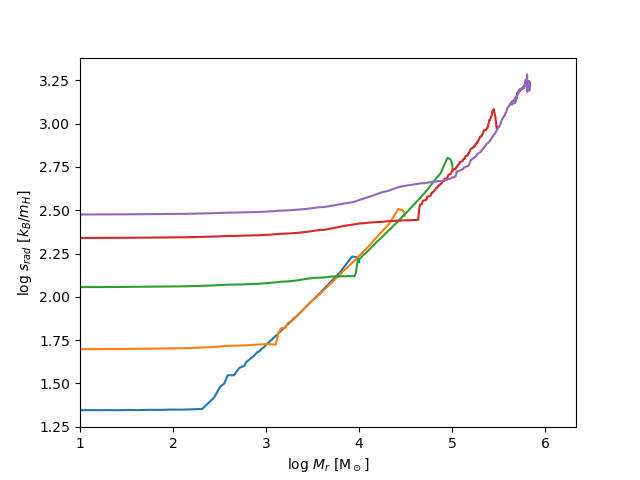}
\caption{Entropy profiles of the model at $Z=3$ Z$_\odot$ and $\dm=10$~\Mpy\
as a function of the Lagrangian coordinate, at different masses.}
\label{fig-entropieZ3_dm10}
\end{center}\end{figure}

\begin{figure}\begin{center}
\includegraphics[width=.5\textwidth]{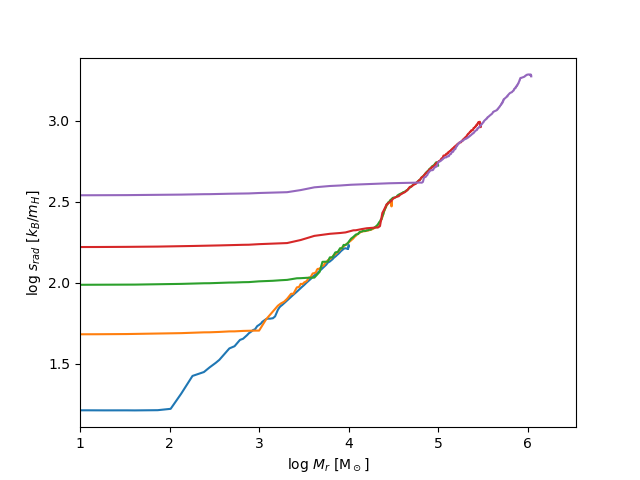}
\caption{Same as Figure~\ref{fig-entropieZ3_dm10} for the model at $Z=3$ Z$_\odot$ and $\dm=100$~\Mpy.}
\label{fig-entropieZ3_dm100}
\end{center}\end{figure}

\begin{figure}\begin{center}
\includegraphics[width=.5\textwidth]{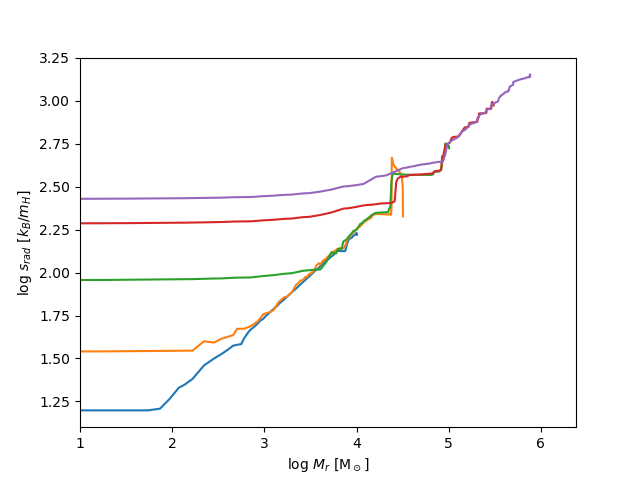}
\caption{Same as Figure~\ref{fig-entropieZ3_dm10} for the model at $Z=3$ Z$_\odot$ and $\dm=1000$~\Mpy.}
\label{fig-entropieZ3_dm1000}
\end{center}\end{figure}

The entropy profiles of the models at $Z=3$ Z$_\odot$
are shown in Figures~\ref{fig-entropieZ3_dm10}-\ref{fig-entropieZ3_dm100}-\ref{fig-entropieZ3_dm1000}.
On each profile, we can distinguish the convective core which is nearly isentropic,
the radiative zone where entropy grows outwards,
and the thin convective envelope where entropy decreases outwards,
because of the non-adiabaticity of convection in the low density outter layers.
We can see that, in the convective core, the entropy is growing due to the heat liberated by H-burning.
In the radiative zone, for accretion rates $\dm>10$~\Mpy, the successive profiles superimpose to each other,
which shows that the evolution is adiabatic in this region of the star:
the entropy losses are negligible in the short evolutionary timescale for such rapid accretion.
In this case, the entropy profiles adjust on hylotropic profiles $s\propto M_r^{1/2}$
\citep{begelman2010,haemmerle2019c,haemmerle2020c}.
Departures from the hylotropic profiles are visible, which reflects the oscillations of the surface described above:
due to the absence of entropy losses, each Lagrangian layer keeps memory of the accretion event.
In particular, for the biggest expansion visible in Figure~\ref{fig-stZ3_dm1000}, at a mass $\sim30\,000$ \Ms,
we find a plateau in the entropy profiles of Figure~\ref{fig-entropieZ3_dm1000}
around the Lagrangian layers $M_r\sim30\,000$~\Ms.
Only for accretion rates $\dm\leq10$~\Mpy\ the evolutionary timescale is long enough
for the losses of entropy due to radiative transfer to be visible in the radiative zone.

\begin{figure}\begin{center}
\includegraphics[width=.5\textwidth]{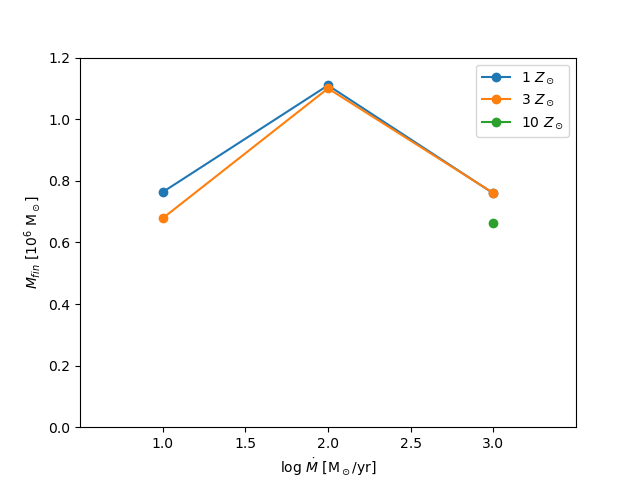}
\caption{Final mass at GR instability as a function of the accretion rate, for the indicated metallicities.}
\label{fig-Mfin}
\end{center}\end{figure}

The final masses at GR instability, obtained with Equation~(\ref{eq-postnewton2}),
are shown for all the models in Table~\ref{tab-mfin} and Figure~\ref{fig-Mfin}.
All the final masses range in the interval $0.6-1.2\times10^6$~\Ms, with a very weak dependence on the metallicity.
For accretion rates $\dm\lesssim100$~\Mpy, the final mass increases as the accretion rate increases.
But for $\dm\gtrsim100$~\Mpy\ the final mass is a decreasing function of the accretion rate.

\begin{table}
\caption{Stellar mass (in $10^6$ \Ms) at the onset of GR instability for the indicated metallicities and accretion rates.}
\label{tab-mfin}
\centering
\begin{tabular}{c|ccc|}
			& 10 \Mpy	& 100 \Mpy	& 1000 \Mpy	\\\hline
$Z=$ Z$_\odot$	&0.763	&1.11	&0.759	\\
$Z=3$ Z$_\odot$	&0.678	&1.10	&0.760	\\
$Z=10$ Z$_\odot$	&		&		&0.663	\\\hline
\end{tabular}
\end{table}

\section{Discussion}
\label{sec-dis}

\subsection{Evolutionary tracks on the HR diagram}

The evolutionary tracks of all the models remain in the red on the HR diagram,
showing no sign of contraction towards the blue until the GR instability, at luminosities $10^{10}-10^{11}$ \Ls.
This is in contrast with the zero-metallicity case, where the tracks start to move to the blue at these luminosities
\citep{hosokawa2013,haemmerle2018a,nandal2024}.
We interpret this difference as the effect of the larger number density of free electrons for higher metallicity,
so that the photosphere is locked on the Hayashi limit towards lower gas densities \citep{haemmerle2019c}.
This fact is key for the problem of the ionising feedback, in particular in the case where other effects
postpone the GR instability (rotation, dark matter; \citealt{haemmerle2021b,haemmerle2024}).
Indeed, in case the star moves to the blue for masses $\gtrsim10^5$~\Ms,
the strong ionising feedback could prevent further accretion \citep{hosokawa2011b,hosokawa2016},
limiting the mass to $\sim10^5$~\Ms.
Here, we see that, for metallicities $\geq$~Z$_\odot$,
the ionising feedback is expected to remain weak up to masses of at least $\sim10^6$~\Ms.

\subsection{Internal structures}

As shown already in \cite{haemmerle2019c} and \cite{haemmerle2020c} for lower metallicities,
when accretion proceeds at rates $\gtrsim10$ \Mpy,
the structure of the star is well approximated by hylotropes \citep{begelman2010},
made of an isentropic core and a radiative envelope where entropy grows outwards like $s\propto M_r^{1/2}$.
At these rates, the evolutionary timescale, set by accretion ($t\sim M/\dm$),
is shorter than the Kelvin-Helmholtz (KH) timescale,
which prevents any loss of entropy by radiative transfer.
The Lagrangian layers of the radiative zone are nevertheless contracting
(Figures~\ref{fig-stZ3_dm10}-\ref{fig-stZ3_dm100}-\ref{fig-stZ3_dm1000}-\ref{fig-stZ1_dm100}).
However this is not a KH contraction, but an adiabatic compression driven by accretion:
the weight of the accreted gas induces an increase in the pressure in the whole star, and so in the density.
As a consequence, the evolution is driven by accretion, all the other processes being too slow to produce any effect.
It implies that the structure of the star is essentially given by the stellar mass, independently of the accretion rate.
On the other hand, another consequence of the adiabaticity of the evolution is that
the successive entropy profiles match each other in the radiative zone:
each Lagrangian layer keeps the entropy it advected at accretion.
It follows that the oscillations of the surface visible in
Figures~\ref{fig-hrZ3}-\ref{fig-stZ3_dm10}-\ref{fig-stZ3_dm100}-\ref{fig-stZ3_dm1000}-\ref{fig-stZ1_dm100}
have an impact on the whole stellar structure, in particular on the size of the convective core.

\begin{figure}\begin{center}
\includegraphics[width=.5\textwidth]{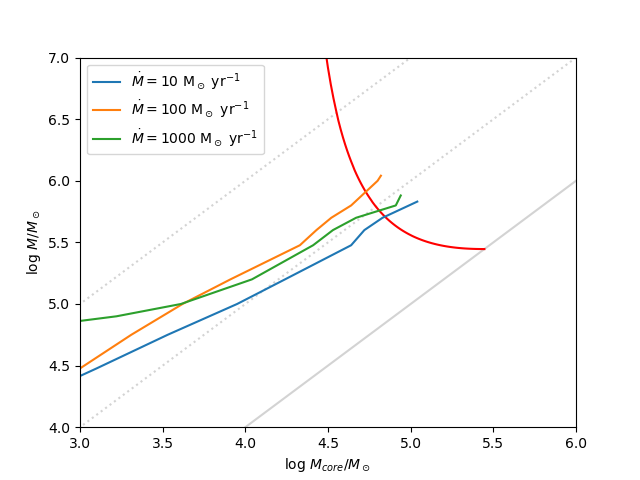}
\caption{$M_{\rm core}-M$ diagram of the models at $Z=3$ Z$_\odot$, for the indicated accretion rates.
The red line is the hylotropic stability limit with $\mu=0.64$ and $T_c=7\times10^7$ K.
The grey diagonals indicate constant fractions $M_{\rm core}/M=1\%-10\%-100\%$.}
\label{fig-hylo}
\end{center}\end{figure}

The evolutionary tracks on the $M_{\rm core}-M$ diagram are shown in Figure~\ref{fig-hylo}
for the models at $Z=3$ Z$_\odot$.
Once the star is supermassive and H-burning has started,
the convective core represents typically 10\% of the total stellar mass.
For a given mass, the model at $\dm=10$ \Mpy\ has a more massive core than the models with larger accretion rates.
This is because of the longer evolutionary timescale in this model, that allows for entropy losses,
so that the layers have the time to contract via a KH contraction,
and to join the core that becomes more massive for a given total mass.
On the other hand, for the models at 100 -- 1000 \Mpy, this effect disappears due to the short evolutionary timescales,
so that the $M_{\rm core}-M$ relation is nearly unique.
Actually, we can see that, for a given mass,
this is the model with the shorter timescale that has the more massive core.
In this regime of extreme accretion, the evolution of the core is impacted
by the advection of the intermediate convective zones, that reflect the oscillations at the stellar surface,
as described above.
In particular, we can see that the oscillations of the surface at $M\sim30\,000$~\Ms\
in the model at $\dm=1000$~\Mpy\ (Figure~\ref{fig-stZ3_dm1000})
translate into oscillations in the $M_{\rm core}-M$ relation when the layers accreted during the oscillations
join the convective core, that is when $M_{\rm core}\sim30\,000$ \Ms.
At this point, the convective core is growing rapidly due to the advection of a series of isentropic layers,
which explains why this model has a larger core than the model at $\dm=100$ \Mpy\ for a given mass.

Finally, we notice that another reason why the 'radiative' zone is prone to transient convective instabilities
is that the adiabatic and radiative gradients are both very close to 1/4 in the Eddington limit:
\begin{equation}
\nabla_{\rm ad}={4-3\beta\over16-12\beta-{3\over2}\beta^2}={1\over4}+\mathcal{O}(\beta^2)
\end{equation}
\begin{eqnarray}
\nabla_{\rm rad}&=&{3\kappa PL_r\over16\pi acT^4GM_r}={1\over4}{1\over1-\beta}{\kappa L_r\over4\pi cGM_r}
={1\over4}{1\over1-\beta}{\dif P_{\rm rad}\over\dif P}	\\
&=&{1\over4}\left(1+{\dln(1-\beta)\over\dln P}\right)
={1\over4}\left(1-\beta{\dln\beta\over\dln P}\right)+\mathcal{O}(\beta^2)
\end{eqnarray}
where we have used the equations of radiative transfer and of hydrostatic equilibrium.
We notice the absence of a $\mathcal{O}(\beta)$ term in the adiabatic gradient.

\subsection{Final masses}

All the models reach the GR instability during H-burning,
so that there is no dark collapse in the sense proposed by \cite{nandal2024}.

All the final masses are $\sim10^6$~\Ms,
with a weak dependence on the accretion rate and on the metallicity.
In particular, the increase in the accretion rate from 100 to 1000 \Mpy\
does not translate into an increase of the final mass as for lower rates.
This fact has already been noticed by \cite{nagele2024} for lower metallicities, but left without explanation.
Here we propose an explanation: because of the adiabaticity of the evolution in this extreme accretion regime,
the stellar structure is determined by the mass only, nearly independently of the accretion rate,
as explained in the previous section, so that we can expect a unique final mass for rates $\gtrsim100$~\Mpy.
Only the differences in the oscillations of the stellar surface,
kept in memory by the gas when it joins the convective core, impact the core's mass.
It leads to a larger core (for given mass) for the model at $\dm=1000$~\Mpy\ compared to the one at $\dm=100$~\Mpy\
(Figure~\ref{fig-hylo}).
As a consequence, the GR instability is reached at a lower mass
for the model at $\dm=1000$~\Mpy\ than for the one at $\dm=100$~\Mpy.

Figure~\ref{fig-hylo} compares the final mass of the models at $Z=3$~Z$_\odot$
with the hylotropic stability limit \citep{haemmerle2020c}.
The hylotropic limit is computed for a mean molecular weight $\mu=0.64$
and a central temperature $T_c=7\times10^7$ K,
typical values for $Z=3$ Z$_\odot$ SMSs in the beginning of H-burning.
We see that the models exceed slightly this analytic limit, but by a maximum of 0.2 dex in $M$.
Thus, the hylotropic limit represents a good approximation for the models.
It illustrates that the larger is the convective core the smaller is the final mass,
explaining why the model at $\dm=1000$~\Mpy\ has a smaller final mass than that at $\dm=100$~\Mpy.

\section{Conclusion}
\label{sec-out}

In the present work, we studied for the first time the properties of SMSs at $Z>$ Z$_\odot$,
focusing on the conditions met in galaxy merger driven direct collapse ($Z=1-10$ Z$_\odot$, $\dm=10-1000$~\Mpy).
We neglected rotation, which will be the topic of the next paper of the series.

We found that, at these accretion rates and metallicities,
SMSs evolve as red supergiant protostars until the GR instability (Figure~\ref{fig-hrZ3}).
The internal structure is made of an isentropic convective core that contains $\sim10\%$ of the total stellar mass,
a 'radiative' region full of small transient convective zones, where entropy grows outwards, following a hylotropic law,
and a thin convective envelope with an entropy decreasing outwards
due to the non-adiabaticity of convection in the low density regions
(Figures~\ref{fig-stZ3_dm10}-\ref{fig-stZ3_dm100}-\ref{fig-stZ3_dm1000}-\ref{fig-stZ1_dm100}-\ref{fig-entropieZ3_dm10}-\ref{fig-entropieZ3_dm100}-\ref{fig-entropieZ3_dm1000}).
For accretion rates $>10$ \Mpy, the evolution in the radiative region is adiabatic
(Figures~\ref{fig-entropieZ3_dm100}-\ref{fig-entropieZ3_dm1000}):
the Lagrangian layers are contracting, however this is not a KH contraction,
but an adiabatic compression driven by accretion.
As a consequence, in this extreme accretion regime, the stellar structure is set essentially by the stellar mass,
independently of the accretion rate.
This fact results in a nearly unique final mass $\sim10^6$~\Ms\ for all the models (Figure~\ref{fig-Mfin}).
Moreover, all the models reach the GR instability during H-burning.


\bibliographystyle{aa}
\bibliography{bib}

\end{document}